\documentclass{aa}
\usepackage{graphicx,natbib}
\begin{document}
\title{Spectroscopic monitoring of the BL Lac object \object{AO 0235+164}
\thanks{Based on observations collected at the European Southern Observatory, Chile
(ESO Programme 71.A-0174),
and on observations made with the Italian Telescopio Nazionale Galileo (TNG) 
operated on the island of La Palma by the Fundaci\'on Galileo Galilei of the INAF 
(Istituto Nazionale di Astrofisica) at the Spanish Observatorio del Roque de los Muchachos 
of the Instituto de Astrofisica de Canarias}}


\author{C.~M.~Raiteri \inst{1}
\and M.~Villata \inst{1}
\and A.~Capetti \inst{1}
\and J.~Heidt \inst{2}
\and M.~Arnaboldi \inst{1}
\and A.~Magazz\`u \inst{3}
}

\offprints{C.\ M.\ Raiteri, \email{raiteri@to.astro.it}}

\institute{INAF, Osservatorio Astronomico di Torino, Via Osservatorio 20,
     10025 Pino Torinese (TO), Italy
\and Landessternwarte Heidelberg-K\"onigstuhl, K\"onigstuhl
D-69117 Heidelberg, Germany
\and INAF-Telescopio Nazionale Galileo, Apartado 565, 38700 Santa Cruz de La  Palma, Spain
}

\date{Received; Accepted;}

\titlerunning{Spectroscopic monitoring of AO 0235+164}

\authorrunning{C.\ M.\ Raiteri et al.}

  \abstract
   {}
   {Spectroscopic monitoring of BL Lac objects is a difficult task
   that nonetheless can provide important information
   on the different components of the active galactic nucleus. 
}
  {We performed optical spectroscopic monitoring of the BL Lac object
  \object{AO 0235+164} ($z=0.94$) with the VLT and TNG telescopes from Aug.\ 2003
  to Dec.\ 2004, during an extended WEBT campaign. The flux of this source
  is both contaminated and absorbed by a foreground galactic
  system at $z=0.524$, the stars of which can act as gravitational
  micro-lenses.
}
  {In this period the object was in an optically faint, though variable state, and
  a broad \ion{Mg}{ii} emission line was visible at all epochs.
  The spectroscopic analysis reveals an overall variation in the \ion{Mg}{ii} line flux
  of a factor 1.9, while the corresponding continuum flux density changed by a factor 4.3.
  Most likely, the photoionising radiation can be identified with the emission
  component that was earlier recognised to be present as a UV--soft-X-ray bump in the source
  spectral energy distribution and that is visible in the optical domain
  only in very faint optical states.
  We estimate an upper limit to the broad line region (BLR) size of a few light months
  from the historical minimum brightness level; from this we infer the maximum amplification
  of the \ion{Mg}{ii} line predicted by the microlensing scenario.}
   {Unless we have strongly overestimated the size of the BLR, only very massive stars could
    significantly magnify the broad \ion{Mg}{ii} emission line, but the time scale of variations 
    due to these (rare) events would be of several years. In contrast, the continuum flux,
    coming from much smaller emission regions in the jet, could be affected by microlensing
    from the more plausible MACHO deflectors, with variability time scales of the order of
    some months.
    }


   \maketitle
%
%
\section{Introduction}

The active galactic nuclei (AGNs) known as BL Lac objects show
extreme variability at all wavelengths from the radio band to the
$\gamma$-rays.
The commonly accepted paradigm foresees that their non-thermal
emission comes from a relativistic plasma jet pointing towards the
observer with a small viewing angle \citep{bla78}.
The consequent relativistic Doppler effect
would be responsible for flux enhancement and contraction of the
intrinsic variability time scales.

These objects by definition show no emission lines or only weak ones,
with equivalent widths not exceeding 5 \AA\ in the rest frame
\citep{sti91}. However, stronger emission lines, in particular broad ones,
have occasionally been detected
in the spectra of a few BL Lacs (e.g.\ AO 0235+164, \citealt{coh87};
\citealt{nil96};
BL Lacertae itself, \citealt{ver95}; \citealt{cor96,cor00}) in optically faint states,
thereby putting their classification into question.

Those broad emission lines recognisable in the spectra of AGN
are produced in the so-called ``broad line region" (BLR),
which is believed to be photoionised by thermal radiation from the accretion disc.
In the case of BL Lac objects, the possibility that photoionisation
can be due to the beamed radiation from the jet is still under debate
\citep{cor00}.

Because of the dramatic flux variability and absence of strong lines,
\citet{ost85} suggested microlensing as a possible
explanation of the BL Lac phenomenon.
According to this interpretation, the moving stars of a galaxy on the line
of sight would act as gravitational microlenses, producing amplification and
the consequent variation in the flux emitted from a compact underlying source.
A microlensing-induced flare would be observed
simultaneously at different wavelengths with time scales that depend on the size of
the regions emitting at the various frequencies.
On the contrary, no amplification of the lines is expected, if the size of the BLR
is of the order of 0.1--1 pc, as classically assumed.
Actually, it was found that the BLR size scales with 
the continuum luminosity and that for some
objects it can be rather compact, of the order of some light days--weeks \citep{wan99,kas00}.
In this case, microlensing would also produce noticeable variations 
in the emission line strengths and profiles \citep{aba02}.
One would thus observe these changes at the same time as
the
continuum ones and would find that the line flux correlates with the continuum flux.
It is worth mentioning that the observed difference between continuum and broad emission line
flux ratios of lensed quasar images has been interpreted as due to microlensing,
leading to an estimate of the BLR size down to 9 light days \citep{met04,way05,kee06}.

In this context, the BL Lac object \object{AO 0235+164} at $z=0.94$ is very interesting.
It is well known for its strong flux variability at all wavelengths.
In the optical band it has shown brightness variations up to $\sim 5$ mag
in 6 months, while radio flux changes up to a factor $\sim 75$ in less than
1 year have been observed. In some cases
simultaneous optical and radio outbursts were observed \citep{bal80,web00,rai01},
with a possible quasi-periodicity of the major radio (and optical) events
\citep{rai01}.

Some observational evidence can be reconciled with the intrinsic
variability scenario for this source only if very extreme relativistic conditions
are assumed: Doppler factors up to $\sim 100$ have been estimated in several papers
\citep{fuj99,kra99,fre00,jor01}. Interstellar scattering would be a viable
extrinsic explanation for the intraday radio variability, at least in some cases,
but it cannot explain variability at the higher frequencies.
The alternative extrinsic scenario, invoking microlensing, appears attractive, since
optical imaging and spectroscopic observations have revealed foreground absorbing systems at
$z=0.524$ and $z=0.851$ \citep[see e.g.][]{coh87,nil96}.
Absorption in excess of Galactic absorption is indeed required to fit the X-ray spectra of
this source \citep[see e.g.][]{mad96,jun04,rai06a}.
The environment of AO 0235+164 is rather complex (see Fig.\ \ref{field}): 
the $z=0.524$ absorber is probably an elliptical galaxy lying 1\farcs 3 
to the east\footnote{Following the notation by \citealt{nil96}, 
it was indicated as G1 in the figure, 
but it is not discernible from the source.}, 
which is very likely interacting with
a Seyfert galaxy 2\arcsec\ south of the source (G2). 
This southern AGN, named ELISA by \citet{rai05}, 
can noticeably affect the source photometry during faint states, especially 
in the blue part of the optical spectrum.
The four galaxies G1--G4 belong to the same $z=0.524$ group \citep{sti93}, 
which might include other 
close objects with still unknown redshifts.

   \begin{figure}
   \caption{Detail of the $R$-band frame of the AO 0235+164 field taken with a 15 s exposure at VLT.
    The image is 1\arcmin $\times$ 1\arcmin\ centred on the source; 
    north is up and east is on the left.
    Following the notation by \citet{nil96}, G1 to G4 are the four galaxies at $z=0.524$;
    the intervening galaxy, G1, is not resolved from the source; the southern AGN, ELISA (G2), is
    instead clearly visible.}
   \label{field}
   \end{figure}

The possibility that microlensing affects the AO 0235+164 emission
was the subject of a number of works in the past
\citep{kay88,sti88,sau92,abr93,kra99,web00}, which each faced the problem
with different techniques (mainly imaging and multiwavelength photometric monitoring).
The results of these studies are contradictory.
According to \citet{sti88}, the presence of a foreground eastern galaxy very close to the 
line of sight,
the fact that \citet{coh87} observed that only mild line flux variations correspond to strong
continuum-flux density changes, and the observation by \citet{bal80}
of contemporaneous optical and radio outbursts, all support a microlensing origin of the
violent variability of this source. Also the new VLBI components emitted after outbursts
are interpreted as images of the lensed core of the source.
In contrast, \citet{kay88} showed how the microlensing scenario is very unlikely on the basis
of several observational data. In particular, he calculated that the separation of the new
VLBI components from the core would imply a non-stellar deflector mass 
of $5.7 \times 10^{5} \, M_{\sun}$.
According to \citet{abr93}, microlensing does not appear to be a likely explanation for
the source properties, while
\citet{kra99} claimed that it can account for the
peculiar variations they observed at three radio frequencies and in the optical band
during a three-week campaign in 1992. Finally, \citet{web00} suggested that the events
in 1975 and 1997, which were characterised by simultaneous optical and radio outbursts,
might be ascribed to microlensing, while others are more likely due to intrinsic causes.

Actually, a careful inspection of the historical optical and radio light curves \citep{rai06b} 
reveals that simultaneity, as well as symmetry of events, are features difficult to verify,
either because of poor sampling (especially lack of optical data on long time intervals due 
to solar conjunctions), or due to the possible overlap of different events.
Even for the best-sampled outburst in 1997--1998, we can only say that there was a 
strong flux increase in both the optical and radio bands, but the details of the outburst 
(exact time of the peak and shape at the various frequencies) largely remain unknown.

Investigation of the microlensing scenario through a comparison between line and continuum
flux variations was performed by
\citet{sau92}, who observed AO 0235+164
in both the $V$ band and a narrow-band filter centred on the \ion{Mg}{ii} line.
He found that the emission-line variability was
comparable to the continuum one and that it was stronger than expected from microlensing.
This conclusion, however, was based on the
models of \citet{nem88}, which assumed large BLR sizes.
More recently, \citet{aba02} has shown that the global amplification of broad emission lines
induced by microlensing events could be noticeable, if the BLR has a size of a few light days,
as found by \citet{wan99} and \citet{kas00} for several AGNs.

In this paper we present an analysis of AO 0235+164 optical spectra taken with the VLT and TNG
telescopes in 2003--2004, when the source was in a faint state.
In this period, a huge multiwavelength observing effort was carried out 
by the Whole Earth Blazar Telescope (WEBT)\footnote{\tt http://www.to.astro.it/blazars/webt/ 
\citep[see e.g.][]{vil04a,vil04b,vil06}}
collaboration to continuously monitor the source emission variability in the optical and radio bands.
Furthermore, three pointings of the XMM-Newton satellite provided information in
the X-ray and UV bands, allowing us to construct the broad-band spectral energy distributions
(SEDs) with simultaneous data \citep{rai05,rai06a,rai06b}.
The analysis of multiepoch SEDs suggested the existence of an extra component besides the
low-energy synchrotron one and the high-energy inverse-Compton one. This extra component,
peaking in the far-UV--soft-X-ray domain, can be explained in terms of an additional
synchrotron component coming from an inner part of the jet with respect to that emitting the
observed radio-to-optical radiation. However, the hypothesis of a thermal component
coming from an accretion disc cannot be ruled out, particularly because of
the possible detection of a strongly redshifted Fe K$\alpha$ line in some of the X-ray spectra,
which should be produced in the inner part of the disc, as discussed by \citet{rai06a}.

\begin{table*}
\centering
\caption{Observation log.}
\begin{tabular}{l c c c c}
\hline
Telescope & Instrument          & Date           & Seeing [\arcsec] & Exp. [s]\\
\hline
VLT4      & FORS2, grism 300V   & Aug.\ 28, 2003 & 0.8              & 900\\
TNG       & DOLORES, grism LR-B & Nov.\ 15, 2003 & 1.3              &1800\\
TNG       & DOLORES, grism LR-B & Jan.\ 17, 2004 & 1.3              &1800\\
TNG       & DOLORES, grism LR-B & Oct.\ 5,  2004 & 1.2              &3600\\
TNG       & DOLORES, grism LR-B & Dec.\ 6,  2004 & 1.4              &3600\\
\hline
\label{log}
\end{tabular}
\end{table*}

\section{Observations and data reduction}

Optical spectra were taken in the period August 2003 -- December 2004
at the 8 m diameter Very Large Telescope (VLT) and in service mode
at the 3.6 m Telescopio Nazionale Galileo (TNG).
The logbook of the observations is shown in Table \ref{log},
where Column 1 reports the telescope, Column 2 the instrument, Column 3 the
observing date, Column 4 the value of seeing derived from the acquisition frame(s), 
and Column 4 the total integration time.
This was obtained in one exposure at VLT, while
three subsequent exposures were acquired at TNG, in order to be able to get rid
of cosmic rays by combining the images.

   \begin{figure*}
   \sidecaption
   \includegraphics[width=12cm]{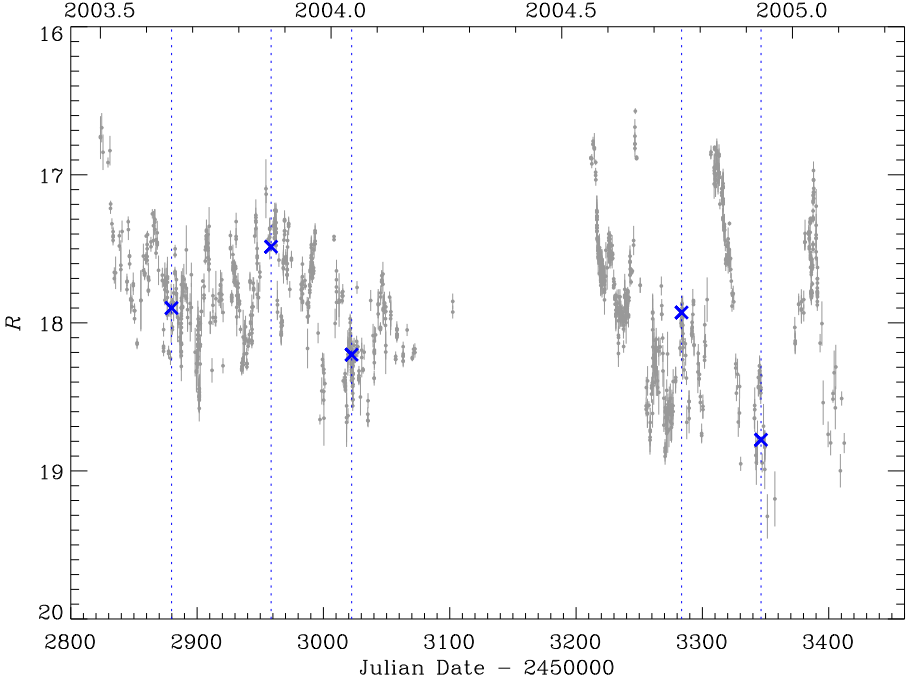}
   \caption{The $R$-band light curve of AO 0235+164 during the 2003--2005
   WEBT campaign (adapted from \citealt{rai06b}). The dates of the spectroscopic observations,
   as well as the corresponding $R$-band values derived from the acquisition images, 
   have been indicated as vertical dotted lines and blue crosses.}
   \label{curva_luce}
   \end{figure*}

All spectra were obtained with a 1\arcsec\ slit aligned along the north-south direction,
so that the light of both the source and ELISA entered the spectrograph.

The data reduction was carried out with the IRAF\footnote{IRAF is distributed by the
National Optical Astronomy Observatories operated by the Association of Universities
for Research in Astronomy, Inc. under cooperative agreement with the 
National Science Foundation} package, following a standard procedure.
Spectra were bias-subtracted and corrected with dome flat-field frames.
Arc-lamp exposures were used for wavelength calibration; 
two-dimensional mapping of pixels into
wavelengths was derived by running the routines {\tt identify}, {\tt reidentify},
and {\tt fitcoords}, which allows correction for geometrical distortions.
Wavelength-calibrated source spectra were then obtained with the
{\tt transform} routine.
Sky-subtracted, one-dimensional source spectra were extracted with {\tt apall}.
A small aperture was used in order to avoid the inclusion of ELISA.
From the VLT frame we could also extract the ELISA spectrum.

Spectra of spectrophotometric standard stars were used for
flux calibration, which was performed with the routines {\tt standard}, {\tt sensfunc},
and {\tt calibrate}.

We checked the correctness of flux calibration using field photometry.
Indeed, for all observations, acquisition frames in at least the $R$ band were taken, 
so that differential photometry was performed with the routine {\tt phot}
with respect to comparison stars in the same field of the source.
The high space resolution (0\farcs 25 per pixel at VLT and 0\farcs 275 per pixel at TNG),
together with the relatively small seeing values (see Table \ref{log}), allowed us to avoid
the inclusion of the ELISA contribution by using an aperture radius of about 1\arcsec.
This is an important point,
since ELISA can significantly affect the photometry of AO 0235+164, 
especially when the source
is in a faint state \citep[see e.g.][]{rai05}.
For the magnitude calibration, we used the photometric sequences by
\citet{smi85} and \citet{fio98}; the values of the source magnitude obtained with respect to the brighest, 
non-saturated comparison stars were averaged, and the uncertainty reported in Table 4 takes into account both
the standard deviation of these values from the mean and the instrumental error.

The results of the photometric measures are reported in Table \ref{res}.
As one can see, the source was always found in faint states, but at different levels,
ranging from $R \sim 17.5$ on November 15, 2003, to
$R \sim 18.8$ on December 6, 2004.
Figure \ref{curva_luce} shows the $R$-band light curve of AO 0235+164 obtained
during the 2003--2005
WEBT campaign (adapted from \citealt{rai06b}).

To check the correctness of the spectroscopic
flux calibration, we transformed the $R$-band magnitudes into flux densities and
compared them with the spectroscopic continua around 6410 \AA.
In order to verify that this procedure also leads to a good calibration
around the \ion{Mg}{ii} line, we derived $V$-band magnitudes 
either from $V$-band frames, when available, or though the mean colour index $V-R =0.76$ 
estimated by \citet{rai05}, obtaining flux densities at $\sim 5450$ \AA.
In a couple of cases, frames in other bands were also available, allowing for a more
detailed comparison between spectroscopic and photometric results.

\subsection{VLT spectra}

The VLT data were acquired on August 28, 2003 under photometric
conditions with sub-arcsecond seeing.
Grism 300V was used, with 11 \AA\ dispersion and 5900 \AA\ central wavelength.
The CCD scale was 0\farcs 25 per pixel, so that
the spectrum of ELISA, the southern AGN located 2\arcsec\ south,
is well-resolved from the source one.
Flux calibration was performed with the spectroscopic standard star LTT 1020, 
which was also used to remove the
prominent telluric absorption feature at 7550--7700 \AA.

\begin{table}
\centering
\caption{Emission lines in the AO 0235+164 spectrum.}
\begin{tabular}{l c c c c}
\hline
Line     & $\lambda_{\rm obs}$ & $z$ & Flux & FWHM\\
         & [\AA]               &     &[$10^{-16} \, \rm erg \, cm^{-2} \, s^{-1}$] & [$\rm km \, s^{-1}$]\\
\hline
$\lambda$2800 \ion{Mg}{ii} & 5431 & 0.940  & 14  & 3500\\
$\lambda$3426 [\ion{Ne}{v}] & 6640 & 0.938  & 1.2 & 700\\
$\lambda$3727 [\ion{O}{ii}] & 7225 & 0.939  & 2 & 600\\
$\lambda$3869 [\ion{Ne}{iii}]& 7503 & 0.939  & 2 & 700\\
$\lambda$4102 H$\delta$   & 7958 & 0.940  & 4 & 3200\\
$\lambda$4340 H$\gamma$   & 8439 & 0.944  & 9 & 3200\\
\hline
$\lambda$3727 [\ion{O}{ii}] & 5682 & 0.524  & 1.3 & 800\\
$\lambda$4861 H$\beta$    & 7407 & 0.524  & 1.8 & 900\\
$\lambda$5007 [\ion{O}{iii}]& 7631 & 0.524  & 1.8 & 900\\
\hline
\label{emission}
\end{tabular}
\end{table}

\begin{figure*}
   \sidecaption
   \includegraphics[width=12cm]{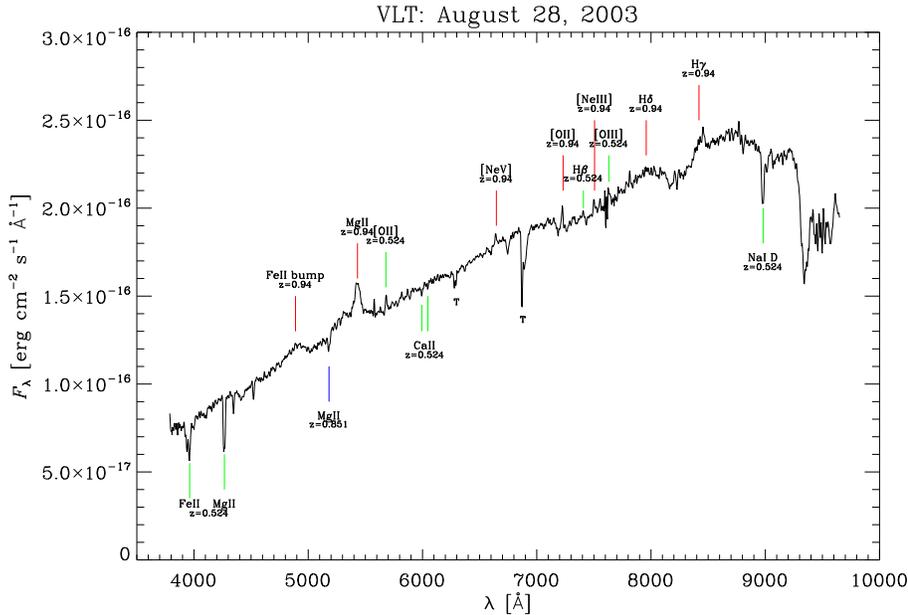}
   \caption{The VLT spectrum of AO 0235+164 acquired on August 28, 2003.
   Several emission features at $z=0.94$, as well as $z=0.524$, are evident, plus
   absorption lines at $z=0.524$ and $z=0.851$.
   Telluric absorption bands are indicated by a ``T" label, and the 7550--7700 \AA\
   telluric feature has been corrected for.}
   \label{VLT_0235}
   \end{figure*}

Figure \ref{VLT_0235} shows the calibrated source spectrum, where the continuum 
is traced  well from about 3800 \AA\ to about 9200 \AA.
The spectrum shows a number of clear emission features at both $z=0.94$ and $z=0.524$
(see Table \ref{emission}). 
In particular, we can distinguish a broad \ion{Mg}{ii} emission line at $\lambda \sim 5430$ \AA,
together with other narrow lines emitted from the source: [\ion{Ne}{v}], [\ion{O}{ii}],
\ion{Ne}{iii}, while the identification of the H$\delta$ and H$\gamma$ lines appears 
more uncertain. The full width half maximum (FWHM) of the \ion{Mg}{ii} emission line is 
$\sim 3500 \rm \, km \, s^{-1}$, a bit larger than the value of 3100 $\rm km \, s^{-1}$ found by \citet{coh87};
on the other hand, the 3200 $\rm km \, s^{-1}$ velocities
we found for the H$\delta$ and H$\gamma$
lines are smaller but consistent with those measured by \citet{nil96}.

The emission lines of [\ion{O}{ii}] and H$\beta$ at $z=0.524$ belong to
the foreground galaxy located at about 1\farcs 3 east of the source.
Two pronounced features, probably due to \ion{Fe}{ii} and \ion{Mg}{ii} absorption
from the same galaxy at $z=0.524$ are evident on the red side of the spectrum, while
on the blue side one can recognise \ion{Na}{i} D absorption at $z=0.524$.
Moreover, the absorption feature at 5183 \AA\ was already ascribed by \citet{coh87}
to \ion{Mg}{ii} at $z=0.851$.

\begin{figure*}
   \sidecaption
   \includegraphics[width=12cm]{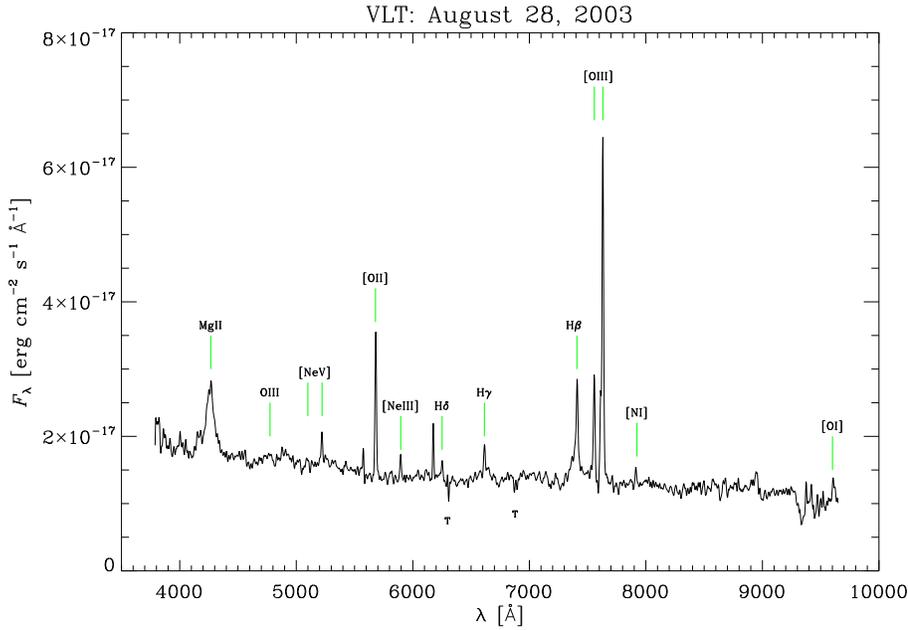}
   \caption{The VLT spectrum of ELISA acquired on August 28, 2003.
   Several emission features are evident;
   telluric absorption bands are indicated by a ``T" label; the 7550--7700 \AA\
   telluric feature has been corrected for.}
   \label{VLT_elisa}
   \end{figure*}

The spectrum of ELISA is shown in Fig.\ \ref{VLT_elisa}. 
Both broad and narrow emission lines are
clearly recognisable. In particular, one can see 
the H$\beta$ and [\ion{O}{iii}] lines already noticed by \citet{coh87}
and by \citet{nil96}. The latter authors also identified the H$\gamma$ line.

All the other features highlighted in Fig.\ \ref{VLT_elisa} are reported for the first time here;
the strong broad \ion{Mg}{ii} and the narrow [\ion{O}{ii}] lines are the most striking features.
Table \ref{elisa} summarises the identified emission lines in the ELISA spectrum, together with the
derived flux.
A double Gaussian fit to the H$\beta$ line to measure both the narrow and the broad components
yields results in agreement with those derived by \citet{nil96}.
We notice that the FWHM of [\ion{O}{ii}] and $\lambda$5007 [\ion{O}{iii}] measured
by \citet{coh87} were 300 and 250 $\rm km \, s^{-1}$, respectively, and the FWHM of the [\ion{O}{iii}] 
lines measured
by \citet{nil96} were 400 $\rm km \, s^{-1}$, somewhat smaller than the ones derived from our spectrum.

\begin{table}
\centering
\caption{Main emission lines in the ELISA spectrum.}
\begin{tabular}{l c c c}
\hline
Line     & $\lambda_{\rm obs}$ & Flux & FWHM\\
         & [\AA]  &[$10^{-17} \, \rm erg \, cm^{-2} \, s^{-1} $] & [$\rm km \, s^{-1}$]\\
\hline
$\lambda$2800 \ion{Mg}{ii} & 4260 & 80 & 4000\\
$\lambda$3426 [\ion{Ne}{v}] & 5220 & 12 & 800\\
$\lambda$3727 [\ion{O}{ii}] & 5682 & 35 & 800\\
$\lambda$3869 [\ion{Ne}{iii}]& 5895 & 5 & 800\\
$\lambda$4102 H$\delta$   & 6252 & 5  & - \\
$\lambda$4340 H$\gamma$   & 6615 & 8  & - \\
$\lambda$4861 H$\beta$ n  & 7410 & 15 & 700\\
$\lambda$4861 H$\beta$ b  & 7410 & 25 &3000\\
$\lambda$4959 [\ion{O}{iii}]& 7556 & 25 & 700\\
$\lambda$5007 [\ion{O}{iii}]& 7630 & 80 & 700\\
$\lambda$5199 [\ion{N}{i}]  & 7915 & 4  & 600\\
$\lambda$6300 [\ion{O}{i}]  & 9606 & 7  & 700\\
\hline
\label{elisa}
\end{tabular}
\end{table}

\begin{table*}
\centering
\caption{$R$-band magnitude, \ion{Mg}{ii} equivalent width and flux, corresponding
continuum flux density, and FWHM of the \ion{Mg}{ii} line
derived from the spectra presented in this paper. EWs and FWHMs are in the observed frame.}
\begin{tabular}{l c c c c c}
\hline
Observation   & $R$  &\ion{Mg}{ii} EW &\ion{Mg}{ii} Flux & Cont.\ Flux Density & FWHM\\
&  [mag]&[\AA] &[$10^{-16} \, \rm erg \, cm^{-2} \, s^{-1} $] & [$10^{-16} \, \rm erg \, cm^{-2} \, s^{-1} \, \AA^{-1}$] &[$\rm km \, s^{-1}$]\\
\hline
VLT, Aug.\ 28, 2003 & $17.90 \pm 0.05$ & $10 \pm 1$     & $14\pm 1$     & $1.4  \pm 0.1$ & 3500\\
TNG, Nov.\ 15, 2003 & $17.49 \pm 0.03$ & $7.1 \pm  0.9$ & $13.8\pm 1.6$ & $1.94 \pm 0.05$ & 3300\\
TNG, Jan.\ 17, 2004 & $18.22 \pm 0.06$ & $12.5 \pm 1.3$ & $11.3\pm 1.1$ & $0.90 \pm 0.06$ & 3300\\
TNG, Oct.\ 5,  2004 & $17.93 \pm 0.02$ & $9.5 \pm 0.8$  & $11.1\pm 0.9$ & $1.17 \pm 0.02$ & 3100\\
TNG, Dec.\ 6,  2004 & $18.79 \pm 0.02$ & $15.7 \pm 1.2$ & $7.2 \pm 0.5$ & $0.45 \pm 0.02$ & 3400\\
\hline
\label{res}
\end{tabular}
\end{table*}

\subsection{TNG spectra}

\begin{figure*}
   \sidecaption
   \includegraphics[width=12cm]{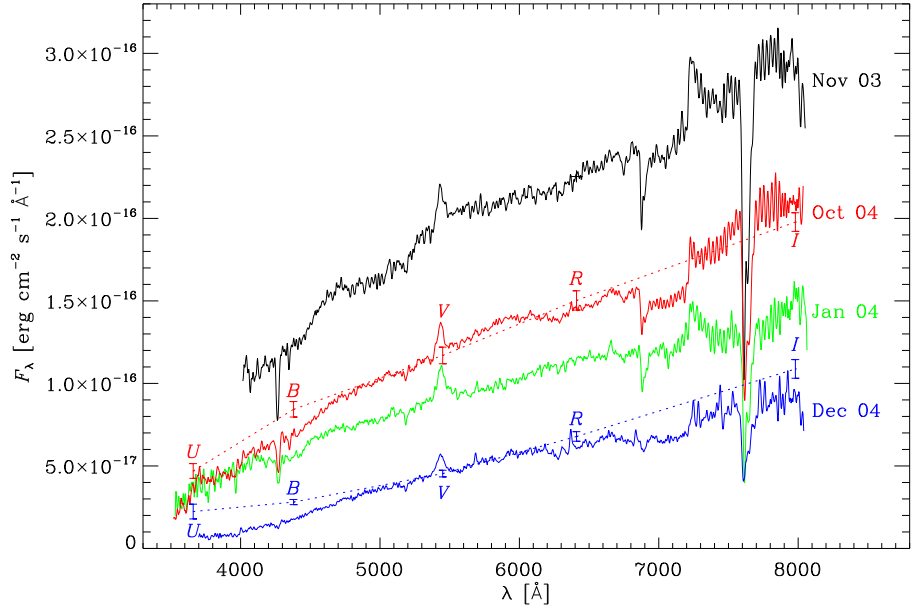}
   \caption{The TNG spectra of AO 0235+164 acquired on November 15, 2003 (black),
    January 17, 2004 (green), October 5, 2004 (red), and December 6, 2004 (blue).
   Flux densities derived from $UBVRI$ photometry in the last two epochs
   are shown to compare the spectroscopic with the photometric shape of the spectra.}
   \label{TNG}
   \end{figure*}

The TNG spectra were obtained with DOLORES and grism LR-B.
The resolution is 11 \AA.
The scale of DOLORES is 0\farcs 275 per pixel, so that the source and ELISA
spectra are resolved also in the TNG frames.
However, the ELISA spectrum cannot be extracted because of the low S/N ratio.
The total integration time (reported in Table \ref{log}) of each observation was obtained with
three subsequent exposures. The three corresponding spectra were reduced and
calibrated separately,
and then summed up in order to increase the S/N ratio.
The measurements of the \ion{Mg}{ii} equivalent width (EW) and flux were performed
on each single exposure; the values reported in Table \ref{res}
are the average values obtained over the exposures, and the uncertainty
corresponds to the standard deviation.
These mean values are in fair agreement with those obtained
by measuring these quantities directly on the summed spectra.
The final summed TNG spectra are shown in Fig.\ \ref{TNG}.

Sky subtraction on October 5 and December 6, 2004 was poor because of CCD charge-dragging problems.
In these two epochs a complete sequence of $UBVRI$ images was taken,
so that it was possible to check the flux calibration at different wavelengths
in order to be sure that the flux around the \ion{Mg}{ii} line was correct.

Figure \ref{TNG} also shows flux densities derived
from $UBVRI$ photometry for these last two epochs. One can notice the deviation from a linear trend
in the bluer part of the spectrum. This deviation is also confirmed by the shape 
of the spectroscopic continuum, even though the photometric data on December 6 give a higher flux
density in $U$ and $B$ bands than do the spectroscopic data.
In any case, the data suggest a blue excess on December 6, when the source was
in the faintest state. This excess would be amplified when correcting for both Galactic and
foreground galaxy absorption, and is likely to be the signature of the low-energy tail of the
extra component suggested by \citet{rai05,rai06a,rai06b}, which has
already been mentioned in the introduction.

\section{Ionising source}

\begin{figure*}
  \sidecaption
   \includegraphics[width=12cm]{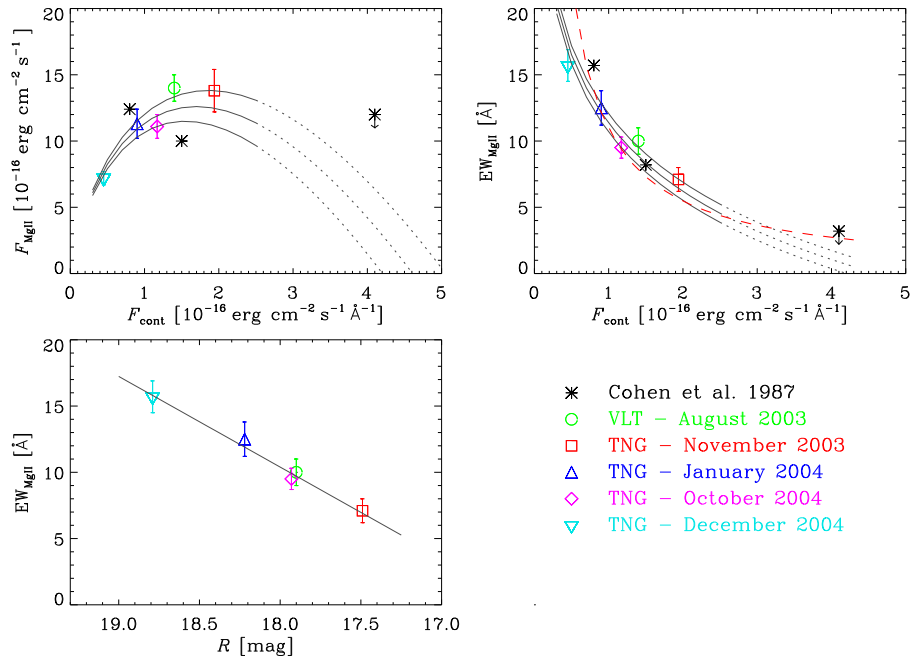}
   \caption{Results of the AO 0235+164 spectroscopic monitoring at the VLT and TNG:
   \ion{Mg}{ii} line flux or equivalent width plotted against continuum flux density
   or $R$-band magnitude; the results by \citet{coh87} are also plotted for comparison.
   See text for further details.}
   \label{all}
   \end{figure*}

Table \ref{res} summarises the results of the spectroscopic analysis on the \ion{Mg}{ii} line.
For each spectrum it reports: the $R$-band magnitude, the \ion{Mg}{ii} line equivalent width
and flux, the corresponding continuum flux density, and the line FWHM.
One can see that the continuum flux density varied by a factor 4.3, while the \ion{Mg}{ii}
flux changed by a factor 1.9. 
The results are also presented in Fig.\ \ref{all}, which shows the 
\ion{Mg}{ii} flux as a function of the continuum flux density,
the \ion{Mg}{ii} equivalent width as a function of the continuum
flux density, and the \ion{Mg}{ii} equivalent width versus the $R$-band magnitude. 
The values obtained by \citet{coh87} are also shown for a comparison.

The main result, confirming the  \citet{coh87} claim, is that the \ion{Mg}{ii} 
flux is variable.
However, its relationship with the
continuum flux density, if any, is not simple.
The \ion{Mg}{ii} equivalent width seems to depend
linearly on the $R$-band magnitude; the grey lines in the various panels represent
this linear relationship:
${\rm EW}=a+b \, R$, with best-fit values $a=-112.76$ and $b=6.84$.
The $R$-band magnitude has been converted into the continuum 
flux density around the \ion{Mg}{ii} line, i.e.\ in the $V$ band, through the $V-R$ index.
The three lines in the first and second panels correspond to the average colour
index and to $1 \, \sigma$ deviations from it according to \citet{rai05}: 
$V-R = 0.76 \pm 0.10$.
For a comparison, in the second panel we also plotted the case where the
\ion{Mg}{ii} line flux is assumed to be constant (${\rm EW} \propto F_{\rm cont}^{-1}$)
and equal to its average
value ($10.94 \times 10^{-16} \, \rm erg \, cm^{-2} \, s^{-1}$; red dashed line).

From Fig.\ \ref{all} one may infer that the line flux increases with the continuum flux density
until a certain brightness level, after which it cannot grow further and may even decrease.
We notice that this first increasing trend depends on just one point, 
but further observations would be required to confirm it.

\citet{cor00} discussed the possible BLR photoionisation source in BL Lacertae.
They analysed eight optical spectra taken in 1995--1997 and measured the H$\alpha$ broad emission line.
The behaviour of its equivalent width as a function of the continuum flux density was 
well-fitted by an $F_{\rm cont}^{-1}$ law.
The interpretation of this lack of response by the line strength
to the continuum variations
in terms of photoionisation by Doppler-boosted synchrotron continuum requires
a high covering fraction of about 35\%.
Although this scenario cannot be ruled out, the authors favoured 
photoionisation by a second continuum source not observable in the optical band.
In the case that it is a hot accretion disc, 
they estimated a temperature $T > 1.2 \times 10^5 \, \rm K$.

For AO 0235+164 this second continuum source can be identified with the UV--soft-X-ray bump
in the SED discussed by \citet{rai06b,rai06a}. From a visual inspection of the SEDs corresponding
to different epochs, one may infer that this component is variable, but its variations do
not appear to be necessarily correlated with the optical ones. 

Thus, for AO 0235+164 we can envisage two scenarios:
i) the BLR is photoionised by the jet emission; the line flux grows with the continuum flux
but only until a kind of saturation level is reached, perhaps corresponding to complete ionisation;
ii) the photoionisation source is another component, producing the UV--soft-X-ray bump in the SED;
it can be detected in the optical domain when the lower-energy synchrotron
component is weak, i.e.\ during very faint optical states.
In this case the line flux is independent of the continuum
flux, since they have different origins, 
and that we see a weaker line in the faintest state is just by chance.
This component can be either emission from a thermal disc or an additional synchrotron component
coming from an inner part of the jet with respect to the one producing the radio--optical emission.
The thermal-disc option may be supported by the possible detection of a
strongly redshifted Fe K$\alpha$ line in the source X-ray spectrum \citep{rai06a}.

Between these scenarios, we think that the second one, i.e.\ photoionisation by a different
component with respect to what produces the optical continuum, is more likely,
since there is observational evidence for its existence.

\section{Size of the BLR}

The size of the BLR in AGNs can be estimated with reverberation mapping techniques, but
relationships have also been derived between the BLR size, $R_{\rm BLR}$,
and the continuum luminosity \citep{wan99,kas00,mcl04}. 
In particular, \citet{mcl04} analysed the dependence of $R_{\rm BLR}$ on the continuum
luminosity at 3000 \AA\ in the source rest frame for high redshift objects ($z > 0.7$).
They found that
$$R_{\rm BLR}=(18.5 \pm 6.6) [ \lambda L_{3000}/10^{37} \, \rm W ]^{(0.62 \pm 0.14)} \, \rm light \, days,$$
where the continuum is supposed to be the ionising radiation of the BLR and can be estimated
around the \ion{Mg}{ii} line.

As discussed in the previous section, the most plausible photoionisation source of the BLR in AO 0235+164
is an extra component, which is visible in the optical frequency range
when the source is faint. Indeed, we recognised a sort of blue excess in the faintest spectrum we analysed 
(taken on December 6, 2004). Hence, if we consider the lowest state of AO 0235+164 ever observed, 
it is {\it plausible} that a substantial fraction of the continuum there comes from the
ionising radiation.

The minimum brightness level of AO 0235+164 in the $V$ band was observed in July 1996: 
$V=20.29 \, \pm \, 0.17$\footnote{From the WEBT archive.}.
Simultaneous data in other bands are not available, but there are $R$-band data
the day before and the day after, indicating a decrease in the source brightness.
Hence, we can assume that their average value, $R=19.89 \pm 0.10$, 
is not far from the actual $R$ value
at the time the $V$-band datum was taken. 
In this case, one gets a colour index $V-R=0.40$, which implies
a rather flatter spectrum than usually found for this source. 
This can be an indication that the higher-energy component was dominant.

If we consider absorption by both the Galaxy and the foreground absorber at $z=0.524$
(1.473 mag in the $V$ band, see \citealt{rai05}), 
we can derive the corresponding de-reddened flux
density at 5450 \AA, $F_\nu=0.108 \pm 0.017 \, \rm mJy$.
Then $\lambda_{\rm e} L_{\lambda_{\rm e}} = 4 \pi d_L^2 \nu F_\nu$,
where $d_L$ is the luminosity distance and $\lambda_{\rm e}$ the emitted wavelength.
Assuming a flat cosmology with
$H_0=71 \, \rm km \, s^{-1} \, Mpc^{-1}$ and $\Omega_{\rm M}=0.27$, $d_L = 6141$ Mpc.
As a result, $\lambda L_{3000} \sim 25.1 \times 10^{37} \rm \, W$.
The \citet{mcl04} equation yields a best fit of $R_{\rm BLR} \sim 136$ light days,
with a range of possible sizes between 56 and 291 light days.

\subsection{Black hole mass}

Recently, \citet{liu06} estimated the black hole mass for AO 0235+164 using several methods.
In particular, they used the equation derived by \citet{mcl04} that applies the virial black hole mass estimate 
$M_{\rm BH} = G^{-1} \, R_{\rm BLR} \, V_{\rm BLR}^2$,
where $V_{\rm BLR}$ is the Keplerian velocity of the line-emitting gas.
The result was a black hole mass of $4.42 \times 10^8 \, M_{\sun}$.
We notice that \citet{liu06} derived the continuum flux from the \citet{coh87}
spectrum of October 1985, when the source was faint, but not at its minimum brightness level, 
and did not correct for Galactic and foreground galaxy absorption.
But, chiefly, they implicitly assumed $V_{\rm BLR}=\rm FWHM_{\ion{Mg}{ii}}$, 
which holds if the BLR gas has a disc-like geometry and if
the angle between the line of sight and the disc axis is $i = 30 \degr$ (see discussion in \citealt{mcl04}).
This is an important point, since the black-hole mass depends on the squared gas velocity.
But if the disc axis coincides with the jet axis, then  
the angle $i$ should be very small for the strongly beamed BL Lacs; 
for AO 0235+164 in particular \citet{fre06} estimated $i = 5 \degr$
by analysing space VLBI maps during a faint state in 2001--2002. 
In this case, however, a disc-like geometry would imply that we should 
systematically see much narrower lines in blazars than in other AGNs (or that blazar black holes are
systematically much bigger).
Hence, the assumption of randomly-oriented orbits for the BLR gas appears more appropriate, 
introducing a factor $\sqrt{3}/2$ into the $V$-FWHM equation.
For a BLR size of 136 light days and an $\rm FWHM_{\ion{Mg}{ii}} \sim 3300 \rm \, km \, s^{-1}$, 
as derived in the previous sections, we then obtain a black hole mass of $\sim 2 \times 10^8 \, M_{\sun}$. 
On the other hand, with the less likely disc geometry and $i \sim 5 \degr$, being 
$V \sim {\rm FWHM}/(2 \sin i)$, we would obtain $M_{\rm BH} \sim 10^{10}\, M_{\sun}$, even if it can be 
intended as an upper limit.

In conclusion, we stress that a black hole mass estimate based on this method suffers from many uncertainties,
in particular those related to the unknown geometry of the BLR, 
so it should be considered with extreme caution.

\section{Microlensing}

To obtain significant amplification of the BLR by microlensing, 
its size should be comparable to the projected radius 
of the Einstein ring of the lens on the source plane $D_{\rm s} \, \theta_{\rm E}$, where
$$\theta_{\rm E}=\sqrt{{{4 G M} \over{c^2}} {{D_{\rm ds}} \over {D_{\rm d} D_{\rm s}}}}$$
is the angular radius of the Einstein ring, 
$M$ the lens mass, and where 
$D_{\rm s}$, $D_{\rm d}$, and $D_{\rm ds}$ are
the angular diameter distances to the source, to the deflector, and between the source and the deflector, 
respectively \citep{ein36}.

With the cosmological assumptions mentioned in Sect.\ 4, $D_{\rm s} = 1632 \, \rm Mpc$,
$D_{\rm d} =  1286 \, \rm Mpc$, and $D_{\rm ds} =  621 \, \rm Mpc$, so that 
the Einstein radius of a 1 $M_{\odot}$ star in the $z=0.524$
galaxy that microlenses emission from AO 0235+164 is about 1.2 $\mu$as.
By comparison, a 136 light-day BLR subtends an angle of about 14 $\mu$as
at the redshift of AO 0235+164. 
The resulting amplification is only about 1\% \citep{aba02,sch92} and
$\sim 8\%$ if the size is 56 light days.
Moreover, the event would last several years, since its duration is given by the 
time the lens takes to cross the projected radius of the Einstein ring on the lens plane:
$t_{\rm var} \sim{ {D_{\rm d} \, \theta_{\rm E}} / {v} } ,$ where $v$ is the transverse 
velocity of the star relative to the source-Earth line, whose typical value
is of the order of $10^3 \, \rm km \, s^{-1}$  \citep{kay86}.
Significant amplification would be obtained in the case of a massive deflector, but
this is expected to be a rare event and would imply even longer time scales. 
Actually, the most likely microdeflectors in galaxies
are thought to be massive compact halo objects (MACHOs) of subsolar mass.

Hence, unless we have strongly overestimated the size of the BLR, we can conclude that we do not
expect to detect variations in the \ion{Mg}{ii} broad emission line due to microlensing
by a star.
On the contrary, since the region emitting the continuum radiation can be significantly 
smaller than the BLR, the possibility remains that microlensing can affect the continuum flux.
To have a typical 2--3 month timescale event with a factor 2--3 flux amplification,
as observed during the big radio outbursts, we would need an $\sim 10^{-4} \, M_{\sun}$
deflector acting on the radiation emitted from a very compact
region of $\sim 0.1$ light days\footnote{ We mention that, according to
\citet{gop91}, very fast variability events can be explained as due to microlensing
if the radiation source is a superluminal blob in the blazar jet.
With this model, \citet{rom95} could explain the strong intraday radio variability of PKS 0537$-$441 
observed in 1993 with lens of $10^{-4}$--$10^{-3} \, M_{\sun}$.}.

More complicated scenarios arise when the optical depth of microlensing increases, so that one cannot
consider isolated stars, but rather their combined effect \citep[see e.g.][]{kay86,sch87,kay89,tor03}.
The investigation of these scenarios goes beyond the aim of this paper, but we note that 
in the case of high optical depth, microlensing produces a sort of flickering rather than
distinct, big events. Moreover, also de-amplification is possible under particular conditions.

The fact that the foreground galaxy of AO 0235+164 is not closely aligned with the blazar 
implies that lens candidates belong to the outer, less populated regions of the galaxy and dark halo, 
where the lensing optical depth is lower.

\section{Conclusions}

We have analysed five optical spectra of the BL Lac object AO 0235+164 taken at the VLT and TNG
in 2003--2004. 
The source was in a faint state, and a broad \ion{Mg}{ii} emission line was clearly
detected in all spectra.
The flux of the line showed an overall variation by a factor 1.9,
while the continuum flux density changed by a factor 4.3.

The observed \ion{Mg}{ii} flux dependence on the continuum flux density may suggest 
that the line flux increases with source brightening but only until a certain saturation level, 
after which it may even decrease.
A further observing effort should be undertaken in order to check this trend.
More likely, the line flux is independent of the optical continuum, 
as found in the case of BL Lacertae \citep{cor00}, and it instead depends on another component.
This can be identified with the one producing the UV--soft-X-ray bump in the source SED.

In this case, the BLR-ionising continuum is most visible in the optical spectral range
when the source is in its faintest optical state. From the historical
$V$-band minimum brightness, we thus estimated the size of the BLR by exploiting its dependence
on the monochromatic luminosity at 3000 \AA\ found by \citet{mcl04}.
The best-fit size is of the order of 4.5 light months, which has however to be taken as an upper limit.

We investigated the possibility of explaining the flux variations we observed in the framework
of a microlensing event due to a star in the foreground galaxy at $z=0.524$. 
Although significant amplification of the continuum flux can also be obtained by subsolar-mass
MACHO-like stars because of the very compact emission region, 
the size of the BLR we derived cannot account for 
a factor-2 line amplification unless we consider microlensing
by a very massive star. Apart from the low probability of encountering such an event,
it would also imply variability time scales of the order of several years, 
in disagreement with the observations.

\begin{acknowledgements}
We thank the referee, Neal Jackson, for useful comments and suggestions.
This work was partly supported by the European Community's Human Potential Programme
under contract HPRN-CT-2002-00321 (ENIGMA).
\end{acknowledgements}

\end{document}